\documentclass{elsart}

\usepackage{graphicx}
\usepackage{amssymb}

\begin{document}

\begin{frontmatter}



\title{The application of slim disk models to ULX:\\the case of M33 X-8}


\author[1]{L. Foschini}
\ead{foschini@bo.iasf.cnr.it}
\author[2,3]{K. Ebisawa}
\author[4]{T. Kawaguchi}
\author[1,8]{N. Cappelluti}
\author[1]{P. Grandi}
\author[1]{G. Malaguti}
\author[5,3]{J. Rodriguez}
\author[3,6]{T.J-L. Courvoisier}
\author[1]{G. Di Cocco}
\author[7]{L.C. Ho}
\author[8]{G.G.C. Palumbo}

\address[1]{INAF/IASF, Sezione di Bologna, Bologna (Italy)}
\address[2]{NASA-GSFC, Greenbelt, MD (USA)}
\address[3]{INTEGRAL Science Data Centre (ISDC), Versoix (Switzerland)}
\address[4]{Optical and Infrared Astronomy Division, NAOJ, Tokyo (Japan)}
\address[5]{CEA Saclay, DSM/DAPNIA/SAp, Gif-sur-Yvette (France)}
\address[6]{Observatoire de Geneve, Sauverny (Switzerland)}
\address[7]{Observatories of the Carnegie Institution of Washington, Pasadena, CA (USA)}
\address[8]{Dipartimento di Astronomia, Universit\`a degli Studi di Bologna, Bologna (Italy)}

\begin{abstract} 
A comparative \emph{XMM-Newton} archival data spectral study of the ultraluminous X-ray source (ULX) M33 X-8  
has been performed by using both the standard disk model and the newly developed slim disk models. The results of this 
analysis tend to confirm the hypothesis that M33 X-8 is an X-ray binary with a stellar-mass black hole 
accreting at super-Eddington rate. 
\end{abstract}


\end{frontmatter}


Since the early observations with the \emph{Einstein} satellite, many ultraluminous X-ray sources (ULX) have been discovered (see, for example, Fabbiano 1989,  Colbert \& Mushotzky 1999, Makishima et al. 2000, Colbert \& Ptak 2002,  Foschini et al. 2002a, Miller \& Colbert 2004). ULX are point-like sources located sufficiently far from the centre of the host galaxy and with X-ray luminosity of about $10^{39-40}$~erg/s in the $0.5-10$~keV energy  band. According to their X-ray spectrum, ULX can be broadly divided into three classes: \\
\emph{Type I:} a thermal component, usually modelled with a multicolour disk (MCD) with an inner disk temperature  $T_{\rm in}\approx 1.0-1.5$~keV, plus a power-law-like excess at high energies.\\
\emph{Type II:} similar to Type I, but with an inner disk temperature $T_{\rm in}\approx 0.2$~keV; these ULX are candidates to host intermediate-mass black holes (e.g. Miller et al. 2003, 2004);\\
\emph{Type III:} a simple power-law model, with a photon index $\Gamma\sim 2$; these ULX could be either background objects 
(e.g. NGC4698 X-1, Foschini et al. 2002b; NGC4168 X-1, Masetti et al. 2003; see Foschini et al. 2002a for a discussion on the
probability to find background AGN in ULX surveys) or stellar-mass black holes in an anomalous very high state (Comptonization-dominated, Kubota et al. 2002).\\ 
It is worth mentioning that some ULX show state transitions, implying also a transition in the types of the present classification (e.g. Kubota et al. 2001a).

The model mainly used to fit the thermal component is based on the standard optically thick and geometrically thin accretion disk by Shakura \& Sunyaev (1973).  The multicolour disk version developed by Mitsuda et al. (1984) is widely applied  in the fit of ULX spectra (\texttt{diskbb} model in \texttt{xspec}). However, this model, together with the power-law model used to fit the hard X-ray excess, presents some problems addressed by several authors (Merloni et al. 2000, Ebisawa et al. 2003, Wang et al. 2004). Wang et al. (2004) proposed a Comptonized version of the multicolour disk (CMCD) to take into account the feedback effects of the disk-corona system and applied this model to some well-known ULX. They found an apparent agreement between the results obtained with the MCD+PL and the CMCD model, despite the known limitations. 

On the other hand, the emission from a number of Galactic black hole candidates (e.g. GRO J$1655-40$, Kubota et al. 2001b,  XTE J$1550-564$, Kubota et al. 2002) cannot be explained in terms of the canonical $L_{\rm disk}\propto T_{\rm in}^4$ relationship between the accretion disk bolometric luminosity and its inner temperature. This behaviour has been interpreted in terms of an  anomalous state which can be represented with an optically and geometrically thick accretion disk, or ``slim disk'' (e.g. Watarai et al. 2000). Since this type of disk can support super-Eddington luminosities, the slim disk model was also proposed to fit some ULX spectra (e.g. Watarai et al. 2001, Kubota et al. 2002, Ebisawa et al. 2003). If this model is correct, some ULX could be understood as stellar-mass black holes with supercritical accretion rates, thus avoiding the introduction of intermediate-mass black holes.

In the present work, we focus on a single ULX -- M33 X-8 (Type I) -- and perform a comparative 
X-ray energy  spectral study by using both the standard disk and the newly developed slim disk 
models. Specifically, for the latter, there are several codes available 
(e.g. Mineshige et al., 2000 without relativistic effects and electron scatterings; 
Watarai et al. 2000, without relativistic effects and a first order approximation for the 
electron scattering), but we adopted the version developed by Kawaguchi (2003), where the 
effects of electron scattering and relativistic correction are included.  Public archival 
data, obtained with \emph{XMM-Newton} observations, are fit by these models in the present study.

M33 X-8 can be considered an object between a well-known binary system (such as Cygnus X-1) 
and a typical ULX (with $L_{\rm X}\approx 10^{39.5}$~erg/s). Therefore, it could be a good 
laboratory to start this type of comparative study.
Since the discovery of M33 X-8 with the \emph{Einstein} satellite (Long et al. 1981), it was 
clear that the source presented peculiar properties ($L_{0.2-4~\rm keV}\approx 10^{39}$~erg/s, 
soft spectrum) and already Trinchieri et al. (1988) suggested that it could be a new type of 
X-ray binary system. Takano et al. (1994) on the basis of \emph{ASCA} data, suggested
that it could be a stellar mass BH ($\approx 10M_{\odot}$), although they were skeptical on the
possibility to find such a peculiar binary system so close to the M33 centre.
Makishima et al. (2000) suggested that X-8 can be considered as a member of ULX population.

A detailed analysis of a set of \emph{XMM-Newton} observations of X-8 has been presented 
by Foschini et al. (2004). The statistically best-fit model is the MCD+PL, in agreement with the
findings of other authors (e.g. Makishima et al. 2000, Takano et al. 1994), but several 
correction factors were needed to infer useful physical parameters. Specifically, for 
the case of a face-on disk, the inferred mass is $(6.2\pm 0.4)M_{\odot}$, which can increase 
up to $(12\pm 1)M_{\odot}$  if the correction factors for Doppler boosting, gravitational 
focusing, and a $\theta=60^{\circ}$ disk inclination, are taken into account. 

\begin{figure}[!t]
\centering
\includegraphics[angle=270,scale=0.25]{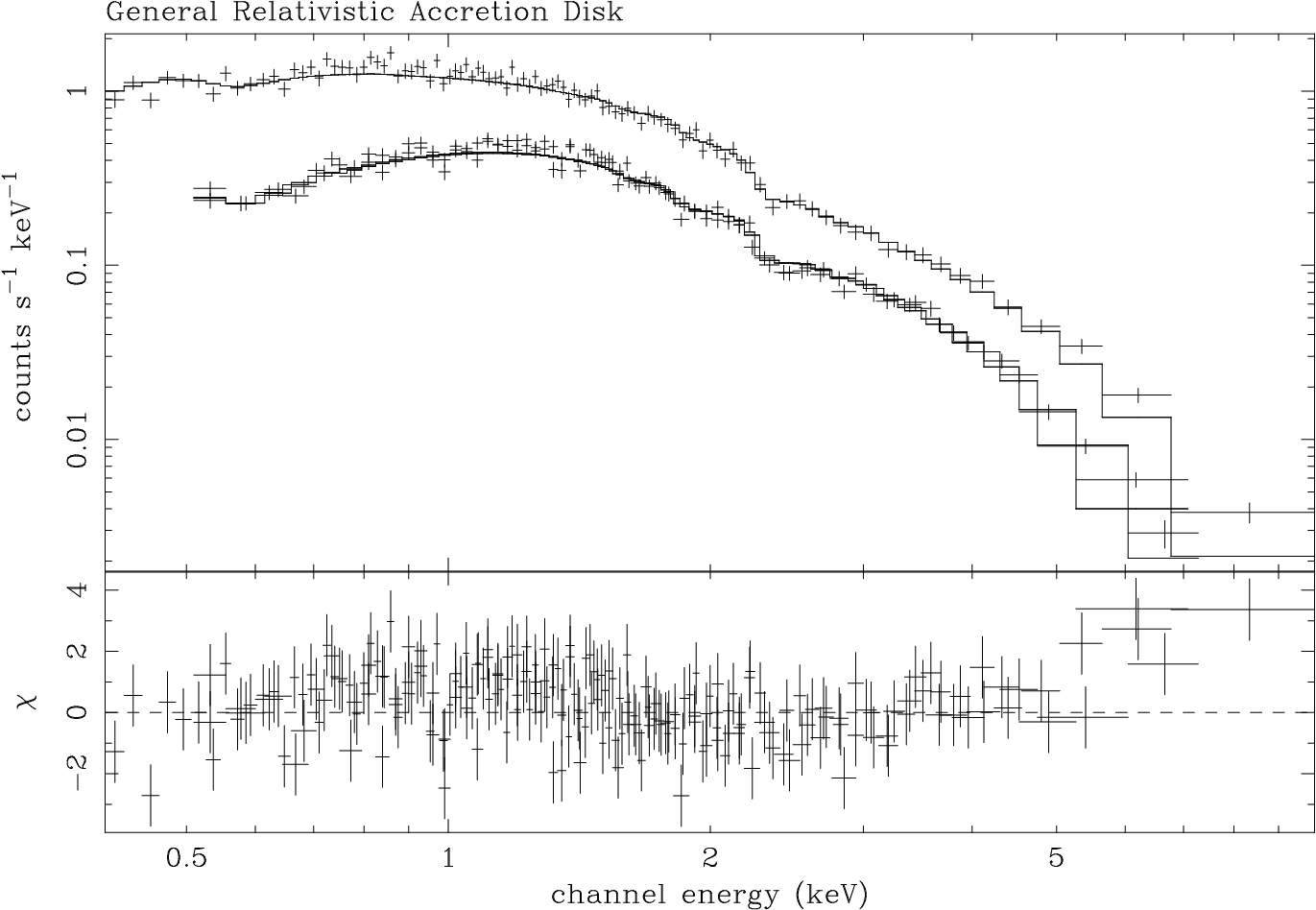}
\includegraphics[angle=270,scale=0.25]{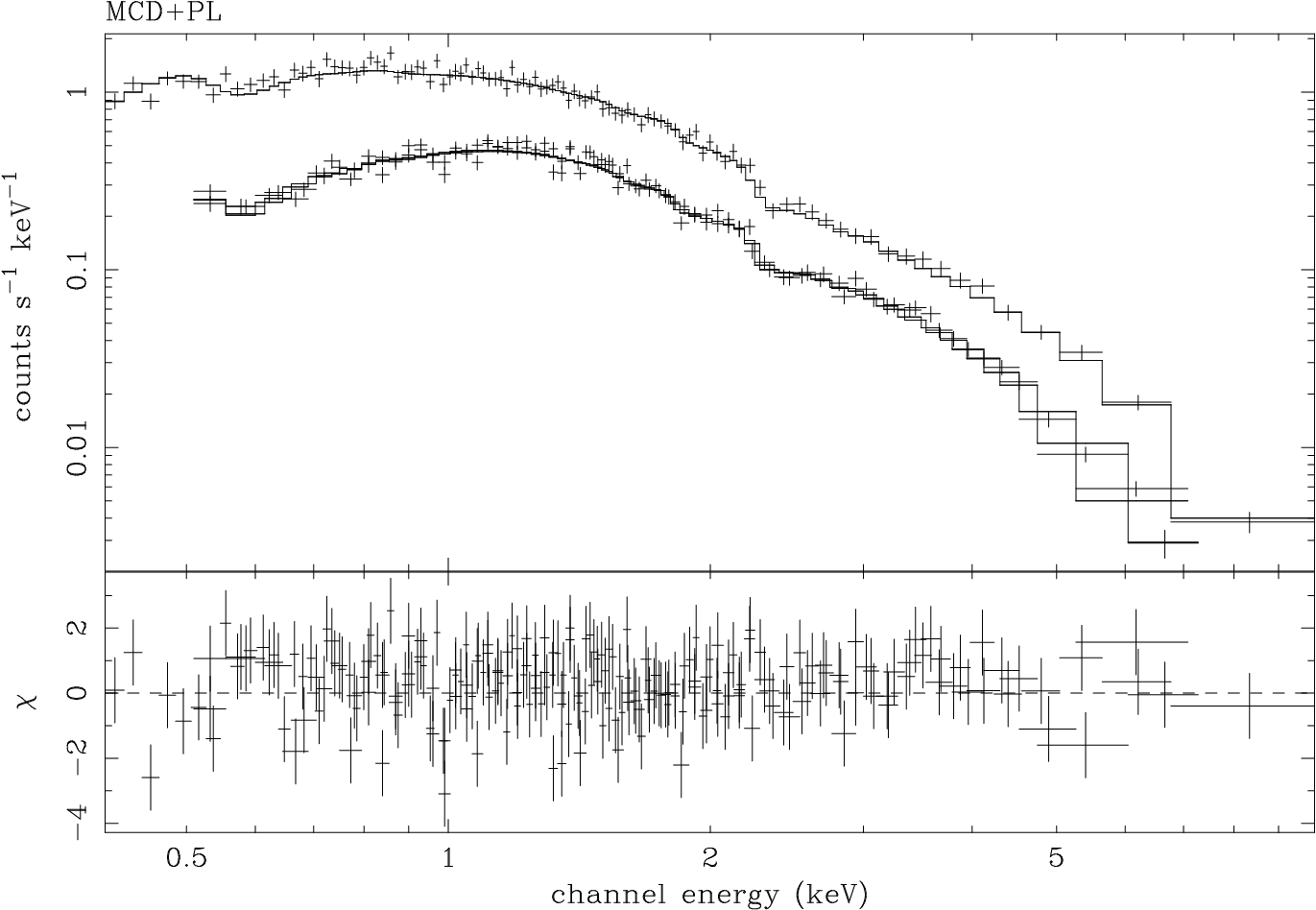}\\
\includegraphics[angle=270,scale=0.25]{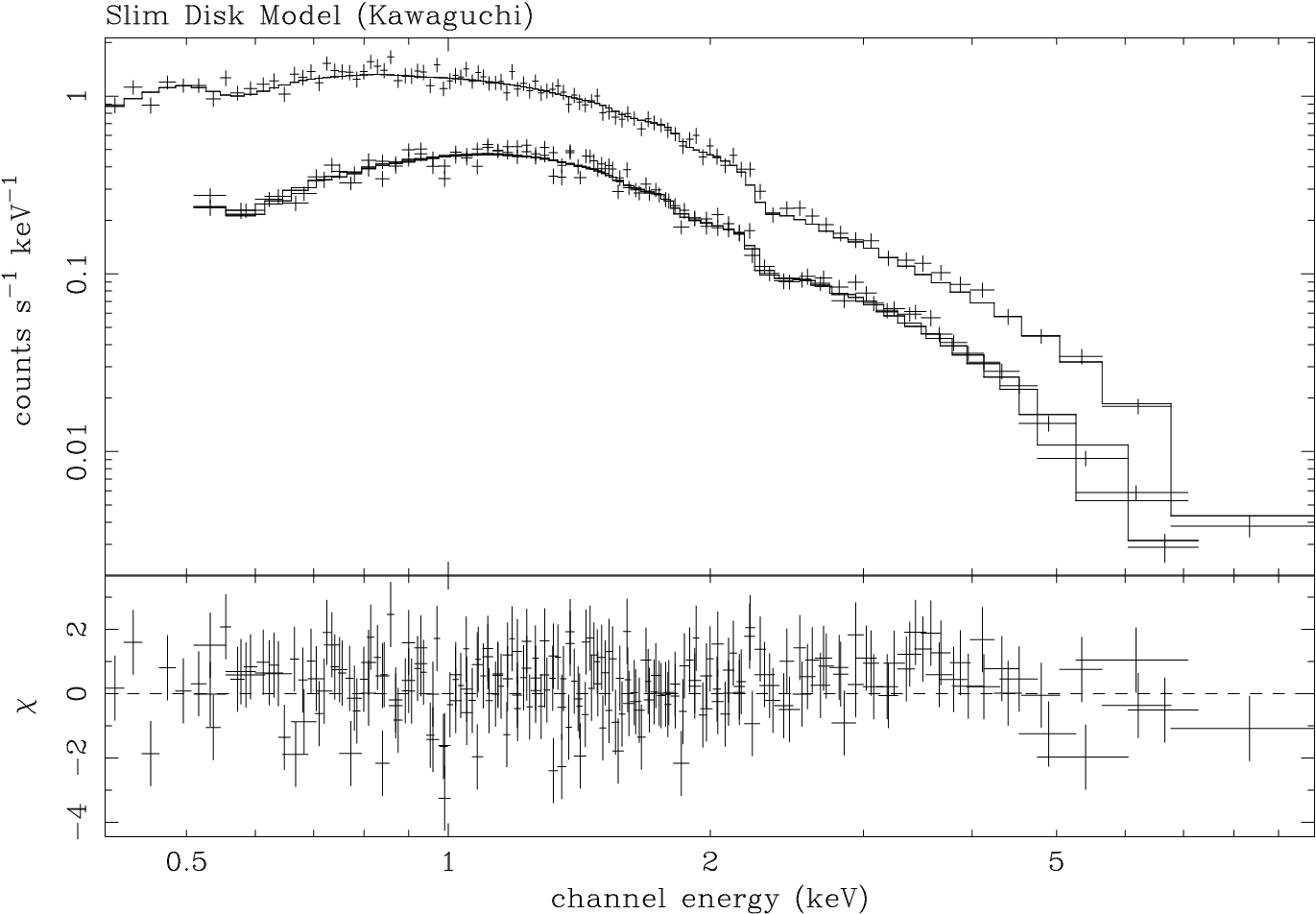}
\includegraphics[angle=270,scale=0.25]{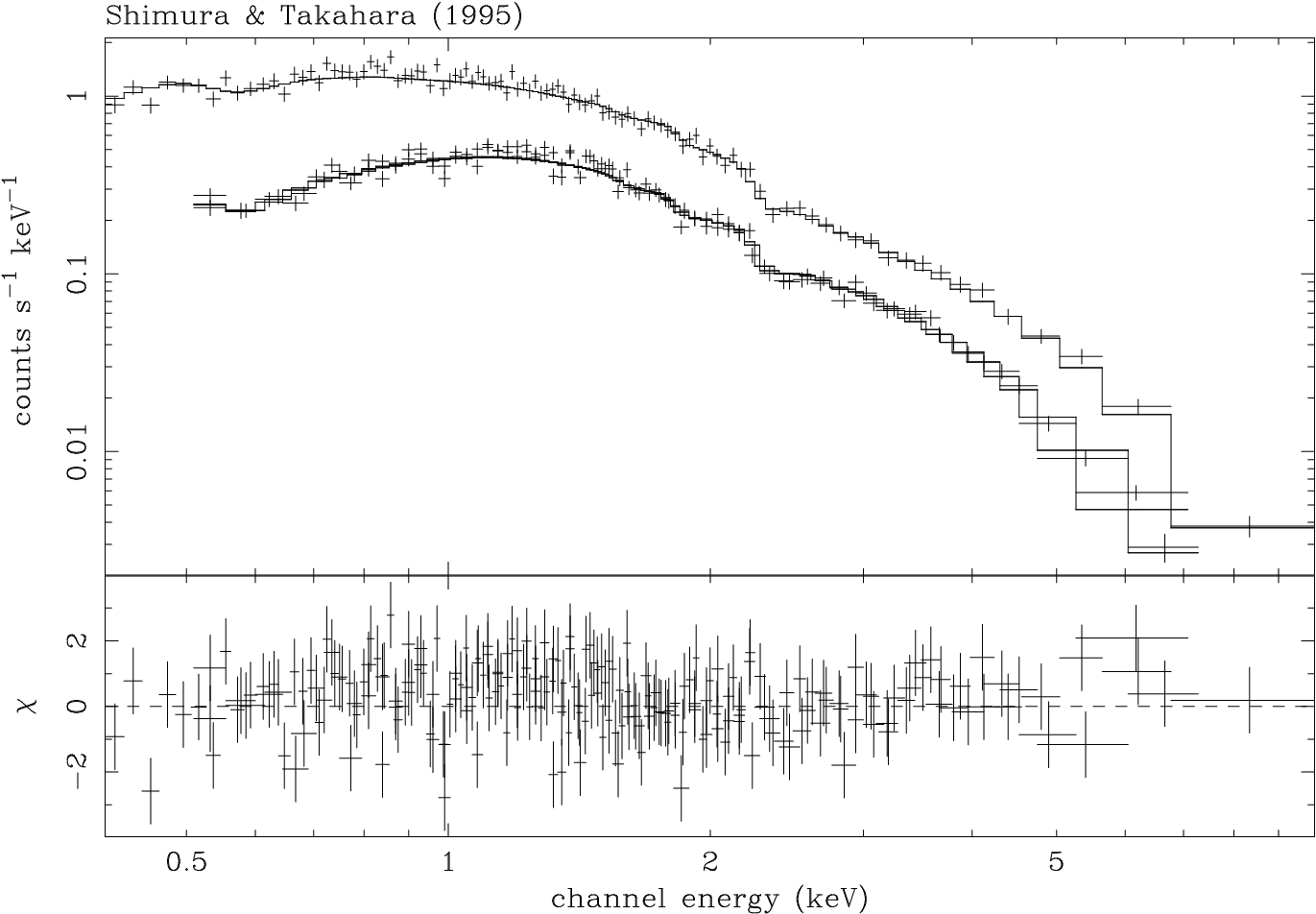}\\
\caption{\emph{XMM-Newton}-EPIC spectra of M33 X-8 (ObsID $0102640101$) and fit residuals to: (\emph{top-left})
general relativistic accretion disk (\texttt{grad} model in \texttt{xspec}); (\emph{top-right}) multicolour
accretion disk (\texttt{diskbb} model) plus a power-law model; (\emph{bottom-left}) slim disk model by
Kawaguchi; (\emph{bottom-right}) Shimura \& Takahara's model.}
\label{spec}
\end{figure}

For the present work, we retrieved from the \emph{XMM-Newton} public archive a set of 
three observations with M33 X-8 on axis (ObsID $0102640101$, $0141980501$, $0141980801$), 
performed on 4 August 2000, 22 January and 12 February 2003, respectively. The data were 
reduced and analyzed in a manner similar to that described in Foschini et al. (2004), except 
for the use of a more recent version of the software (\texttt{XMM-SAS v. 6.1.0} and 
\texttt{HEASoft v 6.0}). Moreover, we analyze here for the first time the 2003 observations, 
now become public.  The spectra were fit in the energy range $0.4-10$~keV for PN and 
$0.5-10$~keV for MOS1 and MOS2 (in this case, the flux was then extrapolated down to $0.4$~keV). 
The width of the energy range was selected to take into account the present status of the 
instrument calibration (Kirsch 2004). The results of the fits are presented in 
Table~\ref{tab:xdata} and Figure~\ref{spec}. 

\begin{table*}[!th]
\caption{Results from the fit of the X-ray data. Models: General relativistic  accretion disk (GRAD;
\texttt{grad} model in \texttt{xspec}), multicolour accretion disk (MCD; \texttt{diskbb} model in
\texttt{xspec}), power-law model (PL), Kawaguchi's (2003)  slim disk model (SDK), Shimura \& Takahara's
(1995) model (ST) with $\alpha$ fixed to $0.1$. Meaning of symbols:  $N_{\rm H}$ absorption columns
[$10^{21}$~cm$^{-2}$], with the Galactic column  $N_{\mathrm H}=5.6\times 10^{20}$ cm$^{-2}$; 
mass $M$ [$M_{\odot}$] (for the MCD+PL model, the mass is calculated from the \texttt{diskbb} model 
normalization); accretion rate  $\dot{M}$ [$L_{\rm Edd}/c^2$]; viscosity parameter $\alpha$; observed flux
$F$ in the $0.4-10$~keV energy band [$10^{-11}$~erg~cm$^{-2}$~s$^{-1}$]; intrinsic luminosity $L$ in the
same energy band [$10^{39}$~erg/s], calculated for $d=795$~kpc;  photon index $\Gamma$; inner disk temperature 
$T_{\rm in}$ [keV]. We considered only the
case of face-on disk. A normalization constant between MOS1/MOS2 and PN in the range $0.83-0.91$
has been used. The uncertainties in the parameters are at the 90\% confidence level.}
\centering
\begin{tabular}{lrrrr}
\hline
Model    & Parameter             & $0102640101$       & $0141980501$        & $0141980801$\\
\hline
GRAD     & $N_{\rm H}$           & $0.56$ (fixed)     & $0.56$ (fixed)      & $0.56$ (fixed)\\
{}       & $M$                   & $5.7\pm 0.2$       & $5.4\pm 0.3$        & $6.5_{-0.2}^{+0.3}$ \\
{}       & $\dot{M}$             & $18.9\pm 0.3$      & $21.1\pm 0.5$       & $16.5\pm 0.3$\\
{}       & $F/L$                 & $1.8/1.5$          & $2.0/1.6$           & $1.5/1.2$\\
{}       & $\tilde{\chi}^2/dof$  & $1.09/763$         & $1.12/361$          & $1.55/650$\\
\hline
SDK      & $N_{\rm H}$           & $1.26\pm 0.06$     & $1.3_{-0.1}^{+0.2}$ & $1.00\pm 0.07$\\
{}       & $M$                   & $6.4_{-0.3}^{+0.6}$& $7_{-2}^{+11}$      & $8\pm 1$ \\
{}       & $\dot{M}$             & $31\pm 3$          & $33_{-21}^{+54}$    & $22_{-3}^{+4}$\\
{}       & $\alpha$              & $<0.005$           & $< 0.01$             & $<0.01$\\
{}       & $F/L$                 & $1.8/1.7$          & $2.1/1.9$           & $1.7/1.5$\\
{}       & $\tilde{\chi}^2/dof$  & $1.00/761$         & $0.99/359$          & $1.15/648$\\
\hline
ST       & $N_{\rm H}$           & $0.78\pm 0.06$      & $0.9\pm 0.1$        & $0.56$ (fixed)\\
{}       & $M$                   & $10.1_{-0.3}^{+0.1}$& $10.1_{-0.7}^{+0.2}$& $8.0\pm 0.3$ \\
{}       & $\dot{M}$             & $11.2_{-0.2}^{+0.3}$& $12.8_{-0.6}^{+0.9}$& $11.8\pm 0.4$\\
{}       & $F/L$                 & $1.8/1.6$           & $2.1/1.8$           & $1.7/1.4$\\
{}       & $\tilde{\chi}^2/dof$  & $1.02/762$          & $1.03/360$          & $1.34/650$\\
\hline
MCD+PL   & $N_{\rm H}$           & $2.2\pm 0.4$          & $2.7_{-0.6}^{+1.1}$   & $2.1_{-0.5}^{+0.3}$\\
{}       & $M$                   & $5.5_{-0.6}^{+0.9}$   & $3.8_{-0.8}^{+0.9}$   & $3.8_{-2.7}^{+42.7}$\\
{}       & $T_{\rm in}$          & $1.21_{-0.09}^{+0.08}$& $1.5_{-0.3}^{+0.2}$   & $0.9_{-0.3}^{+0.7}$\\
{}       & $\Gamma$              & $2.6_{-0.2}^{+0.3}$   & $2.7_{-0.5}^{+0.9}$   & $2.2\pm 0.2$ \\
{}       & $F/L$                 & $1.8/2.1$             & $2.1/2.5$             & $1.7/1.9$\\
{}       & $\tilde{\chi}^2/dof$  & $1.00/760$            & $1.00/358$            & $1.09/647$\\
\hline
\end{tabular}
\label{tab:xdata}
\end{table*}

From a first look at the spectral fits, the model of Shimura \& Takahara (1995) results in a 
varying black hole mass. The GRAD model is not able to fully account for the high-energy 
emission of M33 X-8, thus resulting in a clear excess for $E>5$~keV. The same occurs with the 
MCD model alone (not shown), but by adding a power-law component it is possible to model these 
excesses (see the MCD+PL fit). The slim disk model SDK provides a fit comparable to the MCD+PL, 
with constant mass within the measurement errors. However, it should be noted that 
in order to fit all the data with the MCD+PL models for a constant BH mass, fine-tuning of 
some parameters is needed (spectral boosting factor and inner radius of the disk etc.). In addition, 
we also note that the luminosity of the PL component at $0.4-10$~keV range dominates over the 
disk luminosity. This is not usual for Galactic BH, although two other ULX show similar behaviour
(NGC55 X-7, Stobbart et al. 2004; NGC5204 X-1, Roberts et al. 2005). 

The results presented here are only preliminary results of a comparative study on the 
X-ray spectra of ULX: it is necessary still to improve the models and to extend a more 
detailed analysis on a larger sample of ULX. Particularly, the spectral behaviour vs time 
appears to be the key to understand the nature of ULX. 

\vskip 12pt
\emph{Acknowledgements:} the authors thank an anonymous referee, who helped to improve the
present work.

\vskip 12pt
\textbf{References}\\
\scriptsize{Colbert E.J.M. \& Mushotzky R.F., 1999, ``The Nature of Accreting Black Holes in Nearby Galaxy Nuclei'', ApJ, 519, 89-107\\
Colbert E.J.M. \& Ptak A.F., 2002, ``A Catalog of Candidate Intermediate-Luminosity X-Ray Objects'', ApJS, 143, 25-45\\
Ebisawa K., Zycki P., Kubota A., et al., 2003, ``Accretion Disk Spectra of Ultraluminous X-Ray Sources in Nearby Spiral Galaxies and Galactic Superluminal Jet Sources'', ApJ, 597, 780-797\\
Fabbiano G., 1989, ``X-rays from normal galaxies'', ARA\&A, 27, 87-138\\
Foschini L., Di Cocco G., Ho L.C., et al., 2002a, ``XMM-Newton observations of ultraluminous X-ray sources in nearby galaxies'', A\&A, 392, 817-825\\
Foschini L., Ho L.C., Masetti N. et al., 2002b, ``BL Lac identification for the ultraluminous X-ray source observed in the direction of NGC 4698'', A\&A, 396, 787-792\\
Foschini L., Rodriguez J., Fuchs Y., et al., 2004, ``XMM-Newton observations of the ultraluminous nuclear X-ray source in M 33'', A\&A, 416, 529-536\\
Kawaguchi T., 2003, ``Comptonization in Super-Eddington Accretion Flow and Growth Timescale of Supermassive Black Holes'', ApJ, 593, 69-84\\
Kirsch M., 2004, EPIC status of calibration and data analysis, XMM-SOC-CAL-TN 0018, Issue 2.2\\
Kubota A., Mizuno T., Makishima K., et al., 2001a, ``Discovery of Spectral Transitions from Two Ultraluminous Compact X-Ray Sources in IC 342'', ApJ, 547, L119-L122\\
Kubota A., Makishima K., Ebisawa K., 2001b, ``Observational Evidence for Strong Disk Comptonization in GRO J1655-40'', ApJ, 560, L147-L150\\
Kubota A., Done C., Makishima K., 2002, ``Another interpretation of the power-law-type spectrum of an ultraluminous compact X-ray source in IC 342'', MNRAS, 337, L11-L15\\
Long K.S., D'Odorico S., Charles P.A., Dopita M.A., 1981, ``Observations of the X-ray sources in the nearby SC galaxy M33'', ApJ, 246, L61-L64\\
Makishima K., Kubota A., Mizuno T., et al., 2000, ``The Nature of Ultraluminous Compact X-Ray Sources in Nearby Spiral Galaxies'', ApJ, 535, 632-643\\
Masetti N., Foschini L., Ho L.C., et al., 2003, ``Yet another galaxy identification for an ultraluminous X-ray source'', A\&A, 406, L27-L31\\
Merloni A., Fabian A.C., Ross R.R., 2000, ``On the interpretation of the multicolour disc model for black hole candidates'', MNRAS, 313, 193-197\\
Miller J.M., Fabbiano G., Miller M.C., Fabian A.C., 2003, ``X-Ray Spectroscopic Evidence for Intermediate-Mass Black Holes: Cool Accretion Disks in Two Ultraluminous X-Ray Sources'', ApJ, 585, L37-L40\\
Miller J.M., Fabian A.C., Miller M.C., 2004, ``Revealing a Cool Accretion Disk in the Ultraluminous X-Ray Source M81 X-9 (Holmberg IX X-1): Evidence for an Intermediate-Mass Black Hole'', ApJ, 607, 931-938\\
Miller M.C. \& Colbert E.J.M., 2004, ``Intermediate-mass black holes'', Int. J. Mod. Phys. D, 13, 1-64\\
Mineshige S., Kawaguchi T., Takeuchi M., Hayashida K., 2000, ``Slim-Disk Model for Soft X-Ray Excess and Variability of Narrow-Line Seyfert 1 Galaxies'', PASJ, 52, 499-508\\
Mitsuda K., Inoue H., Koyama K., et al., 1984, ``Energy spectra of low-mass binary X-ray sources observed from TENMA'', PASJ, 36, 741-759\\
Roberts T.P., Warwick R.S., Ward M.J., Goad M.R., Jenkins L.P., 2005, ``XMM-Newton EPIC observations of the ultraluminous X-ray source NGC 5204 X-1'', MNRAS, 357, 1363-1369\\
Shakura N.I. \& Sunyaev R.A., 1973, ``Black holes in binary systems. Observational appearance'', A\&A, 24, 337-355\\
Shimura T. \& Takahara F., 1995, ``On the spectral hardening factor of the X-ray emission from accretion disks in black hole candidates'', ApJ, 445, 780-788\\
Stobbart A.M., Roberts T.P., Warwick R.S., 2004, ``A dipping black hole X-ray binary candidate in NGC 55'', MNRAS, 351, 1063-1070\\
Takano M., Mitsuda K., Fukazawa Y., Nagase F., 1994, ``Properties of M33 X-8, the nuclear source in the nearby spiral galaxy'', ApJ, 436, L47-L50\\
Trinchieri G., Fabbiano G., Peres G., 1988, ``Morphology and spectral characteristics of the X-ray emission of M33'', ApJ, 325, 531-543\\
Wang Q.D., Yao Y., Fukui W., et al., 2004, ``XMM-Newton Spectra of Intermediate-Mass Black Hole Candidates: Application of a Monte Carlo Simulated Model'', ApJ, 609, 113-121\\
Watarai K., Fukue J., Takeuchi M., Mineshinge S., 2000, ``Galactic Black-Hole Candidates Shining at the Eddington Luminosity'', PASJ, 52, 133-141\\
Watarai K., Mizuno T., Mineshinge S., 2001, ``Slim-Disk Model for Ultraluminous X-Ray Sources'', ApJ, 549, L77-L80\\
}
\end{document}